\documentclass[a4paper,12pt]{article}
\usepackage{amsmath}
\usepackage{amssymb}
\usepackage{amsfonts}
\usepackage{amssymb,amsmath}
\usepackage{cancel}
\usepackage[dvips]{lscape,graphicx,epsfig}
\usepackage{fleqn}
\usepackage[footnotesize]{caption}
\usepackage{mathrsfs}

\def\beq{\begin{equation}}
\def\eeq{\end{equation}}
\def\bea{\begin{eqnarray}}
\def\eea{\end{eqnarray}}

\def\<{\left\langle}
\def\>{\right\rangle}

\newcommand{\bc}{\begin{center}}
\newcommand{\ec}{\end{center}}
\newcommand{\bd}{\begin{displaymath}}
\newcommand{\ed}{\end{displaymath}}
\newcommand{\be}{\begin{equation}}
\newcommand{\ee}{\end{equation}}
\newcommand{\ba}{\begin{array}}
\newcommand{\ea}{\end{array}}
\newcommand{\bt}{\begin{tabular}}
\newcommand{\et}{\end{tabular}}

\newcommand{\ds}{\displaystyle}

\begin{document}

\bibliographystyle{OurBibTeX}

\begin{titlepage}

\begin{flushright}
  DCPT-07-188\\
  IPPP-07-94
\end{flushright}
\vspace{2mm}

\begin{center}
{
\sffamily
\LARGE
Theoretical upper bound on the mass of the LSP\\ in the MNSSM\\[8mm]}

S.~Hesselbach$^a$, D.~J.~Miller$^b$, G.~Moortgat-Pick$^a$, 
R.~Nevzorov$^b$\footnote{On leave of absence from the Theory Department, ITEP, Moscow, Russia.}, 
M.~Trusov$^c$ 

\vspace{8mm}

{\small
\it
$^a$ IPPP, University of Durham, Durham, DH1 3LE, U.K.\\
$^b$ Department of Physics and Astronomy, University of Glasgow,\\ 
Glasgow G12 8QQ, U.K.\\ 
$^c$ Theory Department, ITEP, Moscow, 117218, Russia}

\end{center}

\begin{abstract}
\noindent
We study the neutralino sector of the Minimal Non-minimal
Supersymmetric Standard Model (MNSSM) where the $\mu$ problem of the
Minimal Supersymmetric Standard Model (MSSM) is solved without
accompanying problems related with the appearance of domain walls.
In the MNSSM as in the MSSM the lightest neutralino can be the
absolutely stable lightest supersymmetric particle (LSP)
providing a good candidate for the cold dark matter component 
of the Universe. In contrast with the MSSM the allowed range of 
the mass of the lightest neutralino in the MNSSM is limited. We establish 
the theoretical upper bound on the lightest neutralino mass in the 
framework of this model and obtain an approximate solution for this 
mass.
\end{abstract}

\end{titlepage}
\newpage
\setcounter{footnote}{0}

\section{Introduction}

The analysis of fluctuations in the cosmic microwave background (CMB)
using recent WMAP satellite data \cite{Spergel:2003cb} and other 
precise measurements \cite{2} indicate that about 22\%-25\% of the energy 
density of the Universe exists in the form of stable non--baryonic, 
non--luminous matter, so called dark matter \cite{Bertone:2004pz}. 
Although the microscopic composition of dark matter remains a mystery 
it is clear that it can not consist of any elementary particles which 
have been discovered so far. Thus the existence of dark matter is 
the strongest piece of evidence for physics beyond the Standard Model 
(SM) of electroweak interactions.

The minimal supersymmetric (SUSY) standard model (MSSM) is the best 
motivated extension of the SM nowadays. Within the MSSM the quadratic
divergences, which destabilise the scale hierarchy, are
cancelled \cite{5} and the gauge coupling unification can be naturally 
achieved \cite{6}. If R--parity is conserved the lightest supersymmetric 
particle (LSP) in the MSSM is absolutely stable and can play the role 
of dark matter \cite{7}. In most supersymmetric scenarios the LSP is 
the lightest neutralino. Since neutralinos are heavy weakly interacting 
particles they explain well the large scale structure of the Universe 
\cite{Primack:2002th} and can provide the correct relic abundance of dark 
matter if their masses are of the order of the electroweak (EW) scale \cite{7}.   
  
Despite these successes the MSSM suffers from the so-called $\mu$--problem. 
Namely, the MSSM superpotential contains only one bilinear term 
$\mu (\hat{H}_d \epsilon  \hat{H}_u)$. In order to get the correct pattern 
of electroweak symmetry breaking, the parameter $\mu$ is required to be of 
the order of the electroweak scale. While the corresponding coupling is 
stable under quantum corrections, it is rather difficult (although 
possible \cite{9}) to explain within Grand Unified theories (GUTs) or 
supergravity (SUGRA) why the dimensionful parameter $\mu$ should be so 
much smaller than the Planck or Grand Unification scale. 

In the Next--to--Minimal Supersymmetric Standard Model (NMSSM) 
\cite{11}--\cite{12}, which contains an additional SM singlet
superfield $\hat{S}$, a $Z_3$ symmetry forbids any bilinear terms in the 
superpotential allowing the interaction of $\hat{S}$ with the
Higgs doublets $\hat{H}_u$ and $\hat{H}_d$:
$\lambda \hat{S}(\hat{H}_d\epsilon\hat{H}_u)$. 
At the EW scale the superfield $\hat{S}$ gets a non-zero vacuum 
expectation value ($\langle S \rangle =s/\sqrt{2}$)
generating automatically an effective $\mu$-term 
($\mu_{eff}=\lambda s/\sqrt{2}$) of the required size.
There is a number of phenomenological reasons which make the NMSSM and its
modifications 
quite attractive. First of all fine tuning which is needed to evade
the LEP~II Higgs 
mass bounds is less severe within SUSY models with an extra singlet
field as compared 
with the MSSM \cite{19}. The upper bound on the lightest Higgs boson
mass in the singlet  
extensions of the MSSM was studied recently in \cite{20}. The spectrum
of Higgs bosons 
in the considered models depends on how strongly the Peccei--Quinn
symmetry is broken  
in these models \cite{Panagiotakopoulos:2000wp}--\cite{20a}. Another
nice feature is  
related with the electroweak baryogenesis which is easier to achieve
in SUSY models  
with an extra singlet field than in the MSSM due to additional terms in
the tree--level  
potential \cite{21}--\cite{Menon:2004wv}. Recently
SUSY models with extra singlet fields including their 
implications for dark matter and neutralino collider  
searches \cite{24} and neutrino physics \cite{25}
have been studied.

However, the NMSSM suffers from a domain wall problem in the early
Universe which can be avoided in the Minimal Non--minimal
Supersymmetric Standard Model (MNSSM) as will be discussed in section 2.1\,.
In this letter we consider the neutralino sector of the MNSSM.
We concentrate on the mass 
of the lightest neutralino because it can be absolutely stable and
therefore may play  
the role of the cold dark matter. We establish a theoretical upper
bound on the lightest  
neutralino mass in the MNSSM which depends rather strongly on the
parameters of the 
considered model. In the allowed part of the parameter space the mass
of the lightest  
neutralino does not exceed $80-85\,\mbox{GeV}$. This permits to
distinguish the MNSSM 
from the MSSM and other SUSY models with an extra singlet superfield at
future colliders. 
We also find an approximate solution for the lightest neutralino mass.
It will enable us to estimate the mass of this particle if charginos
and Higgs bosons  
are discovered in the nearest future. The article is organised as follows.
In the next section we define the MNSSM in more detail.
In section 3 we
examine the allowed range of the lightest
neutralino mass  
in the MNSSM and in section 4 we obtain an approximate solution for its mass. 
Our results are summarised in section 5.

\section{The MNSSM}

\subsection{Superpotential}

As already mentioned the NMSSM itself is not without problems.
The vacuum expectation values of the Higgs fields break the $Z_3$
symmetry in the NMSSM.
This leads to the formation of domain walls in the early Universe \cite{13} 
which create unacceptably large anisotropies in the cosmic microwave
background  radiation \cite{14}. In an attempt to break the $Z_3$ symmetry
operators suppressed  
by powers of the Planck scale could be introduced. But these operators give 
rise to a quadratically divergent tadpole contribution, which destabilises the 
mass hierarchy \cite{15}. Dangerous operators can be eliminated if an
invariance under $Z_2^R$ or $Z_{5}^R$ symmetries is imposed 
\cite{Panagiotakopoulos:1998yw}--\cite{Panagiotakopoulos:1999ah}. The
linear term 
$\Lambda\hat{S}$ in the superpotential which is induced in this case
by high order
operators is too small to upset the mass hierarchy but large enough
to prevent the 
appearance of domain walls.
The corresponding simplest extension of the MSSM is the 
Minimal Non--minimal Supersymmetric Standard Model (MNSSM)
\cite{Panagiotakopoulos:2000wp}, \cite{Menon:2004wv},
\cite{Panagiotakopoulos:1999ah}--\cite{Dedes:2000jp}.
The superpotential of the MNSSM can be written as
\be
W_{MNSSM}=\lambda \hat{S}(\hat{H}_d \epsilon \hat{H}_u)+\xi \hat{S}
+W_{MSSM}(\mu=0)\,.
\label{2}
\ee

\subsection{Neutralino and chargino sectors}

The neutralino sector in SUSY models is formed by the superpartners of
the neutral gauge and Higgs bosons.
Since the sector responsible for electroweak symmetry breaking in the
MNSSM contains an extra singlet field the 
neutralino sector of this model includes one extra component besides the four MSSM ones.
This is an additional Higgsino $\tilde{S}$ (singlino) which is the fermion component of the singlet 
superfield $\hat{S}$. After the breakdown of gauge symmetry the Higgsino mass terms in the MNSSM Lagrangian are 
induced by the trilinear interaction $\lambda \hat{S}(\hat{H}_d \hat{H}_u)$ in the superpotential (\ref{2}). As a result
their values are determined by the coupling $\lambda$ and the vacuum expectation values of Higgs fields. 
The gaugino masses are set by $M_1$ and $M_2$ which are the $SU(2)$ and $U(1)_Y$ soft gaugino mass parameters
that break global supersymmetry. In supergravity models with uniform gaugino masses at the Grand Unification 
scale the renormalisation group flow yields a relationship between $M_1$ and $M_2$ at the EW scale, i.e. 
$ M_1\simeq 0.5 M_2$ . The mixing between gauginos and Higgsinos is proportional to the corresponding 
gauge coupling and the vacuum expectation value of the scalar partner of the considered Higgsino. 
Thus after the electroweak symmetry breaking the superpartners of the electromagnetically neutral 
components of the Higgsino doublets $\tilde{H}^0_d$ and
$\tilde{H}^0_u$, of the singlino $\tilde{S}$ as well as 
the electromagnetically neutral $SU(2)$ and $U(1)_Y$ gauginos ($\tilde{W}_3$ and $\tilde{B}$) mix forming 
a $5\times 5$ neutralino mass matrix which in the interaction basis 
$(\tilde{B},\,\tilde{W}_3,\,\tilde{H}^0_d,\,\tilde{H}^0_u,\,\tilde{S})$ reads
\be
M_{\tilde{\chi}^0}=
\left(
\ba{ccccc}
M_1                  & 0                  & -M_Z s_W c_{\beta}   & M_Z s_W s_{\beta}  & 0 \\[2mm]
0                    & M_2                & M_Z c_W c_{\beta}    & -M_Z c_W s_{\beta} & 0 \\[2mm]
-M_Z s_W c_{\beta}   & M_Z c_W c_{\beta}  &  0                   & -\mu_{eff}         & -\ds\frac{\lambda v}{\sqrt{2}} s_{\beta} \\[2mm]
M_Z s_W s_{\beta}    & -M_Z c_W s_{\beta} & -\mu_{eff}           &  0                 & -\ds\frac{\lambda v}{\sqrt{2}} c_{\beta} \\[2mm]
0                    & 0                  & -\ds\frac{\lambda v}{\sqrt{2}} s_{\beta}  & -\ds\frac{\lambda v}{\sqrt{2}} c_{\beta} & 0       
\ea
\right)\,,
\label{1}
\ee
where $s_W=\sin\theta_W$, $c_W=\cos\theta_W$, $s_{\beta}=\sin\beta$, $c_{\beta}=\cos\beta$ and $\mu_{eff}=\ds\frac{\lambda s}{\sqrt{2}}$.
Here we introduce $\tan\beta=v_2/v_1$ and
$v=\sqrt{v_1^2+v_2^2}=246\,\mbox{GeV}$, where $s$, $v_1$ and $v_2$ are
the vacuum expectation 
values of $S$, $H_d$ and $H_u$, respectively.

The top--left $4\times 4$ block of the mass matrix (\ref{1}) contains the neutralino mass matrix of the MSSM where the parameter $\mu$ is 
replaced by $\mu_{eff}$. From Eq.~(\ref{1}) one can easily see that the neutralino spectrum in the MNSSM may be parametrised in terms of
\be
\lambda\,,\qquad \mu_{eff}\,,\qquad \tan\beta\,, \qquad M_1\,,\qquad M_2\,.
\label{3}
\ee 
The mass parameters $M_2$ and $\mu_{eff}$ also define the masses of
the charginos, the superpartners of the charged gauge and Higgs bosons.
Since the SM singlet superfield $\hat{S}$ is electromagnetically neutral it
does not contribute any extra particles to the chargino
spectrum. Consequently the chargino mass matrix and its eigenvalues remain the same as in the MSSM, namely
\be
\ba{rcl}
m^2_{\chi^{\pm}_{1,\,2}}&=&\ds\frac{1}{2}\biggl[M_2^2+\mu_{eff}^2+2 M^2_{W}\pm\biggl.\\[2mm]
&&\biggr.\qquad\qquad\qquad\sqrt{(M_2^2+\mu^2_{eff}+2M^2_{W})^2-4(M_2\mu_{eff}-M^2_{W}\sin 2\beta)^2}
\biggr]\,.
\ea
\label{4}
\ee
Unsuccessful LEP searches for SUSY particles including data collected at $\sqrt{s}$ between $90\,\mbox{GeV}$ and 
$209\,\mbox{GeV}$ set a $95\%$ CL lower limit on the chargino mass of about $100\,\mbox{GeV}$ \cite{53}. This lower bound 
constrains the parameter space of the MNSSM restricting the absolute values of the effective $\mu$-term and $M_2$
from below, i.e. $|M_2|$, $|\mu_{eff}|\ge 90-100\,\mbox{GeV}$.

\section{Upper bound on the mass of lightest neutralino}

Theoretical restrictions on the masses of the neutralinos cannot be
established using directly the neutralino mass matrix because its eigenvalues
can in general be complex.
In order to find appropriate bounds on these masses it is much
more convenient to consider the matrix $M_{\tilde{\chi}^0} M^{\dagger}_{\tilde{\chi}^0}$ whose eigenvalues are
positive definite and equal to the absolute values of the neutralino
masses squared. In the field basis
$(\tilde{B},\,\tilde{W}_3,\,\tilde{H}^0_d,\,\tilde{H}^0_u,\,\tilde{S})$ the hermitian matrix 
$M_{\tilde{\chi}^0} M^{\dagger}_{\tilde{\chi}^0}$ takes the form:
\be
\left(
\ba{ccccc}
|M_1|^2+M_Z^2 s_W^2  & -M_Z^2 c_W s_W       & -M_Z s_W A^{*}                  & M_Z s_W B^{*}                   & 0 \\[2mm]
-M_Z^2 c_W s_W       & |M_2|^2+M_Z^2 c_W^2  & M_Z c_W C^{*}                   & -M_Z c_W D^{*}                  & 0 \\[2mm]
-M_Z s_W A           & M_Z c_W C            & |\mu_{eff}|^2+\rho^2            & (\nu^2-M_Z^2)c_{\beta}s_{\beta} & \nu^{*}\mu_{eff} c_{\beta}\\[2mm]
M_Z s_W B            & -M_Z c_W D           & (\nu^2-M_Z^2)c_{\beta}s_{\beta} & |\mu_{eff}|^2+\sigma^2          & \nu^{*}\mu_{eff} s_{\beta}\\[2mm]
0                    & 0                    & \nu\mu^{*}_{eff} c_{\beta}          &  \nu\mu^{*}_{eff} s_{\beta}               & |\nu|^2
\ea
\right),
\label{5}
\ee
where
$$
\ba{rclrcl}
\rho^2&=&M_Z^2c_{\beta}^2+\nu^2s_{\beta}^2\,,\qquad\qquad&\sigma^2&=&M_Z^2s_{\beta}^2+\nu^2c_{\beta}^2\,,\qquad 
\nu=\ds\frac{\lambda v}{\sqrt{2}}\,,\\[2mm]
A&=&M_1^{*}c_{\beta}+\mu_{eff}s_{\beta}\,,\qquad\qquad& C&=&M_2^{*}c_{\beta}+\mu_{eff}s_{\beta}\,,\\[2mm]
B&=&M_1^{*}s_{\beta}+\mu_{eff}c_{\beta}\,,\qquad\qquad& D&=&M_2^{*}s_{\beta}+\mu_{eff}c_{\beta}\,.
\ea
$$
Since the minimal eigenvalue of any hermitian matrix is less than its smallest diagonal element 
at least one neutralino in the MNSSM is always light, because 
the mass of the lightest neutralino is limited from above by the
bottom--right diagonal
entry of matrix (\ref{5}), i.e. $|m_{\chi^0_1}|\lesssim
|\nu|$ \cite{Hesselbach:2007ta}. Therefore in contrast to the MSSM
the lightest neutralino in the MNSSM remains light even when the SUSY
breaking scale tends to infinity.

However, the obtained theoretical bound on the lightest neutralino mass can be improved significantly.
In order to get a more stringent limit on $|m_{\chi^0_1}|$ one can
perform an unitary transformation of matrix (\ref{5})
so that $M_{\tilde{\chi}^0}M^{\dagger}_{\tilde{\chi}^0}\to U
M_{\tilde{\chi}^0}M^{\dagger}_{\tilde{\chi}^0} U^{\dagger}$, where
\be
U=\left(
\ba{ccccc}
1  & 0  & 0          & 0          & 0 \\[2mm]
0  & 1  & 0          & 0          & 0 \\[2mm]
0  & 0  & -s_{\beta} & c_{\beta}  & 0 \\[2mm]
0  & 0  & c_{\beta}  & s_{\beta}  & 0 \\[2mm]
0  & 0  & 0          & 0          & 1 
\ea
\right).
\label{6}
\ee
As a result we get
\be
\left(
\ba{ccccc}
|M_1|^2+M_Z^2 s_W^2  & -M_Z^2 c_W s_W       & M_Z s_W \tilde{A}^{*}                 & -M_Z s_W \tilde{B}^{*}                  & 0 \\[2mm]
-M_Z^2 c_W s_W       & |M_2|^2+M_Z^2 c_W^2  & -M_Z c_W \tilde{C}^{*}                
& M_Z c_W \tilde{D}^{*}                   & 0 \\[2mm]
M_Z s_W \tilde{A}    & -M_Z c_W \tilde{C}   & |\mu_{eff}|^2+\tilde{\rho}^2          & \ds\frac{(\nu^2-M_Z^2)}{2}\sin4\beta    & 0 \\[2mm]
-M_Z s_W \tilde{B}   & M_Z c_W \tilde{D}    & \ds\frac{(\nu^2-M_Z^2)}{2}\sin4\beta  & |\mu_{eff}|^2+\tilde{\sigma}^2          & \nu^{*}\mu_{eff}\\[2mm]
0                    & 0                    & 0                                     &  \nu\mu^{*}_{eff}                       & |\nu|^2
\ea
\right),
\label{7}
\ee
where
$$
\ba{rclrcl}
\tilde{\rho}^2&=&M_Z^2\sin^2 2\beta+|\nu|^2\cos^2 2\beta\,,\qquad\qquad&\tilde{\sigma}^2&=&M_Z^2\cos^2 2\beta+|\nu|^2\sin^2 2\beta\,,\\[2mm]
\tilde{A}&=&M_1^{*}\sin 2\beta+\mu_{eff}\,,\qquad\qquad& \tilde{B}&=&M_1^{*}\cos 2\beta\,,\\[2mm]
\tilde{C}&=&M_2^{*}\sin 2\beta+\mu_{eff}\,,\qquad\qquad& \tilde{D}&=&M_2^{*}\cos 2\beta\,.
\ea
$$
Since we can always choose the field basis in such a way 
that the bottom-right $2\times 2$ block of the mass matrix (\ref{7})
becomes diagonal its two eigenvalues also restrict the mass interval 
of the lightest neutralino.
In particular, the absolute value of the lightest neutralino mass
squared has to be always less than 
or equal to the minimal eigenvalue $\mu^2_0$
of this submatrix, i.e. 
\be
|m_{\chi^0_1}|^2\lesssim \mu_{0}^2=\ds\frac{1}{2}\biggl[|\mu_{eff}|^2+\tilde{\sigma}^2+|\nu|^2-
\sqrt{\biggl(|\mu_{eff}|^2+\tilde{\sigma}^2+|\nu|^2\biggr)^2-4|\nu|^2\tilde{\sigma}^2}\biggr]\,.
\label{8}
\ee 
The value of $\mu_0$ decreases with increasing $|\mu_{eff}|$, hence reaching
its maximum value, i.e.
$\mu_0^2=\min\{\tilde{\sigma}^2,\, |\nu|^2 \}$, when
$\mu_{eff}\to 0$. Taking the LEP bound on $\mu_{eff}$ into account 
and also the theoretical upper bound
on the Yukawa coupling $\lambda$ which is caused by the requirement
that the perturbation theory is valid up to 
the Grand Unification scale, requiring $\lambda<0.7$, we find that
$\mu_0^2<0.8 M_Z^2$, i.e.\ $|m_{\chi^0_1}| < 80 - 85$~GeV.

Here it is worth to notice that at large values of the effective $\mu$--term the theoretical restriction on
$|m_{\chi^0_1}|$ (\ref{8}) tends to zero independently of the value of
$\lambda$. Indeed, for $|\mu_{eff}|^2 \gg M_Z^2$ we have
\be
|m_{\chi^0_1}|^2\lesssim \ds\frac{|\nu|^2\tilde{\sigma}^2}{\biggl(|\mu_{eff}|^2+\tilde{\sigma}^2+|\nu|^2\biggr)}\,.
\label{9}
\ee
Thus in the considered limit the lightest neutralino mass is
significantly smaller than $M_Z$ even for large values of
$\lambda \sim 0.7$.

\section{Approximate solution}

\subsection{Characteristic equation}

The masses of the lightest neutralino can be computed numerically by
solving the characteristic equation
$\mbox{det}\left(M_{\tilde{\chi}^0}-\varkappa I\right)=0$. In the MNSSM the corresponding characteristic polynomial
has degree $5$ because the neutralino spectrum is described by a
$5\times 5$ mass matrix. After a few simple algebraic  
transformations we get
\be
\ba{c}
\mbox{det}\left(M_{\tilde{\chi}^0}-\varkappa I\right)=
\biggl(M_1M_2-(M_1+M_2)\varkappa+\varkappa^2\biggr)\biggl(\varkappa^3-(\mu_{eff}^2+\nu^2)\varkappa+\\[2mm]
+\nu^2\mu_{eff}\sin 2\beta\biggr)+M_Z^2\biggl(\tilde{M}-\varkappa\biggr)\biggl(\varkappa^2+
\mu_{eff}\sin 2\beta \varkappa-\nu^2\biggr)=0\,,
\ea
\label{10}
\ee
where $\tilde{M}=M_1 c_W^2 + M_2 s^2_W$. Although one can find a
numerical solution of Eq.~(\ref{10}) for each set of
the parameters (\ref{3})
it is worth to derive either an exact or approximate solution of the
characteristic equation (\ref{10})
to explore the dependence of the lightest neutralino mass on these parameters.
Unfortunately, in the general case the exact solution of this equation
is very complicated.
But in the limit when one of the eigenvalues of the mass matrix
(\ref{1}) goes to zero one can obtain an approximate 
solution of Eq.~(\ref{10}).
Indeed, if $\varkappa\to 0$ we can ignore all higher order terms with
respect to $\varkappa$ in the
characteristic equation keeping only the term which is proportional to
$\varkappa$ and the $\varkappa$--independent one.
The application of this method is justified in the MNSSM because
the mass of the lightest neutralino is limited from above and the
upper bound on $|m_{\chi^0_1}|$ tends to zero with increasing
$|\mu_{eff}|$ or decreasing $\lambda$, as argued in the previous section.
 Actually one can easily check that for a reasonable choice of the
 parameters ($\mu_{eff}, M_2\gtrsim 200\,\mbox{GeV}$,  
$\lambda=0.1-0.7$, $\tan\beta=3-20$ and $M_2\sim 0.5 M_1$) the
lightest neutralino mass is always 
significantly less than the mass of the second lightest one.
Therefore, we can expect that the 
approximate solution obtained in this way would describe the exact one
with high accuracy in a large part of the parameter space. 

However, if we proceed in that way it would mean that we would allow
only one neutralino to be light.
Then the lightest neutralino mass will be consistently described
if the four other neutralino states are considerably heavier than $M_Z$.
One can expect that at least three neutralino states which correspond
to the superpartners of the neutral components of the Higgs doublets
and of the neutral $SU(2)$ gauge boson satisfy this requirement,
because $|M_2|< M_Z$ and $|\mu_{eff}|< M_Z$ are ruled out by chargino 
searches at LEP.
But
the mass of the neutralino state which is predominantly the superpartner
of the $U(1)_Y$ gauge boson is set by $M_1$ which 
may have a value below $M_Z$. If there are two light states in the
neutralino spectrum then the coefficient in front 
of the linear term with respect to $\varkappa$ in Eq.~(\ref{10}) may
be relatively small.
In this case the term which is proportional to $\varkappa^2$ should be
taken into account as well 
in order to 
obtain a suitable approximate solution for the mass of the lightest
and second lightest neutralino.
The inclusion of the quadratic term 
improves the agreement between the numerical and approximate solutions
even when the second lightest neutralino is 
heavier than $M_Z$. Omitting all higher order terms involving
$\varkappa^n$ with $n>2$ in the characteristic equation we find 
\be
A \varkappa^2-B\varkappa+C=0,
\label{11}
\ee
where
\be
\,A = 1+\ds\frac{\nu^2-M_Z^2}{\mu_{eff}^2+\nu^2}\frac{\mu_{eff}\sin 2\beta}{M_1+M_2}+
\ds\frac{M_Z^2}{\mu_{eff}^2+\nu^2}\frac{\tilde{M}}{M_1+M_2}\,,
\label{12}
\ee
\be
\ba{rcl}
B&=&\ds\frac{M_1 M_2}{M_1+M_2}+\biggl(\frac{\nu^2}{\mu_{eff}^2+\nu^2}-
\ds\frac{M_Z^2}{\mu_{eff}^2+\nu^2}\frac{\tilde{M}}{M_1+M_2}\biggr)
\mu_{eff}\sin 2\beta-\\[4mm]
&&-\ds\frac{M_Z^2\nu^2}{(M_1+M_2)(\mu_{eff}^2+\nu^2)}\,,
\ea
\label{13}
\ee
\be
\,C=\ds\frac{\nu^2}{\mu_{eff}^2+\nu^2}\biggl(\frac{M_1M_2}{M_1+M_2}\mu_{eff}\sin 2\beta-
\frac{\tilde{M}}{M_1+M_2}M_Z^2\biggr)\,.
\label{14}
\ee
In order to reduce the characteristic equation (\ref{10}) to
Eq.~(\ref{11}) we have divided both parts of this equation by 
$(\mu_{eff}^2+\nu^2)(M_1+M_2)$.
One can simplify Eq.~(\ref{11}) even further taking into account 
that the second and last terms in Eq.~(\ref{12}) can be neglected
since they are much smaller than unity in most of  
the phenomenologically allowed part of the MNSSM parameter space.
Then the mass of the lightest neutralino can be approximated by
\be
|m_{\chi^0_1}|=\mbox{Min}\left\{\frac{1}{2}\biggl|B-\sqrt{B^2-4C}\biggr|,\,
\frac{1}{2}\biggl|B+\sqrt{B^2-4C}\biggr|\right\}
\,.
\label{15}
\ee

\subsection{Numerical results and discussion}

In Fig.~1 (a)--(c) we plot the numerical and approximate solutions for the lightest neutralino mass as a function of $\mu_{eff}$, $M_2$ 
and $\tan\beta$. For simplicity we assume that all parameters (\ref{3}) appearing in the neutralino mass matrix are real.
We also choose $M_1=0.5 M_2$ and $\lambda=0.7$ which is the largest possible value of $\lambda$ that does not spoil the
validity of perturbation theory up to the GUT scale. From Fig.~1 (a)--(b) it becomes clear that $|m_{\chi^0_1}|$
attains its maximum at certain values of $M_2$ and $\mu_{eff}$. The corresponding maximum value of $|m_{\chi^0_1}|$ is
always less than the upper bound on the lightest neutralino mass
derived in the previous section.

\setlength{\unitlength}{1cm}
\begin{figure}[tp]
\centering
\begin{picture}(9,21)
\put(0,14.5){\epsfig{file=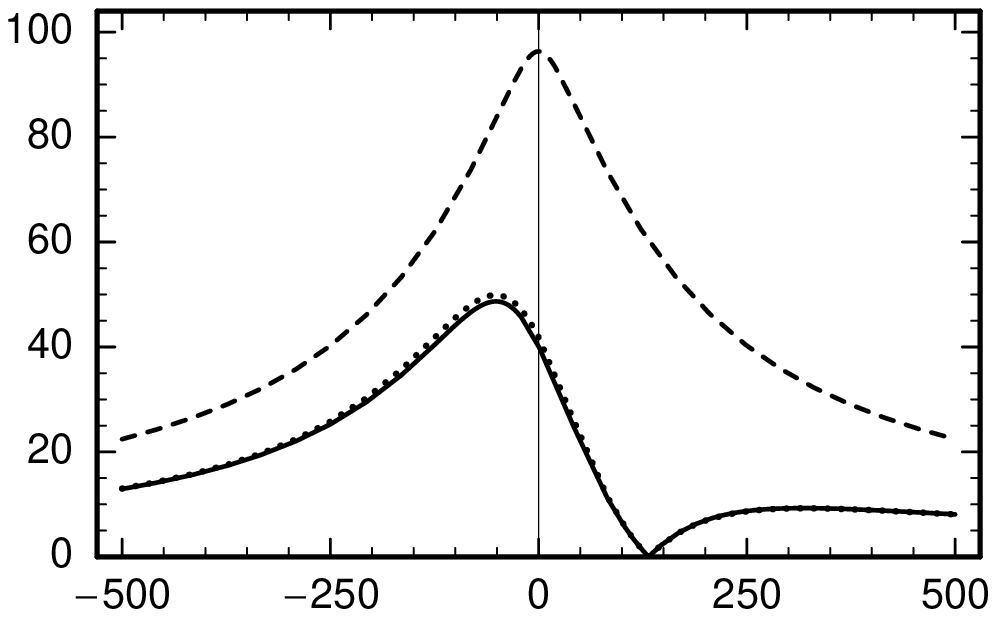,scale=0.9}}
\put(1.2,19.5){(a)}
\put(0.1,20.4){$|m_{\tilde{\chi}^0_1}|$ [GeV]}
\put(4,14.2){$\mu_{eff}$ [GeV]}
\put(0,7.5){\epsfig{file=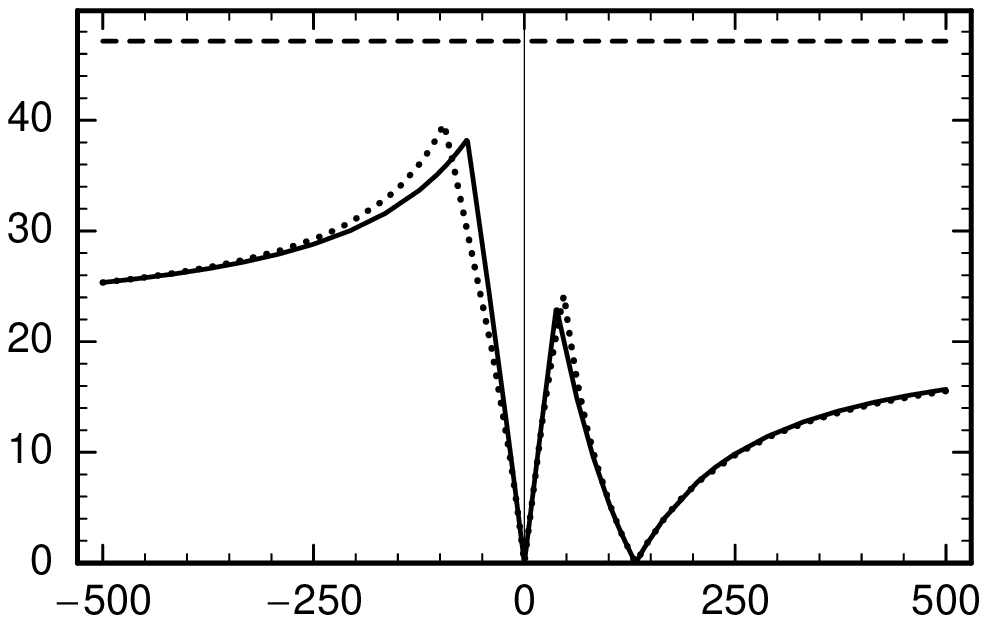,scale=0.9}}
\put(1.2,12.3){(b)}
\put(0.1,13.4){$|m_{\tilde{\chi}^0_1}|$ [GeV]}
\put(4,7.2){$M_2$ [GeV]}
\put(0,0.5){\epsfig{file=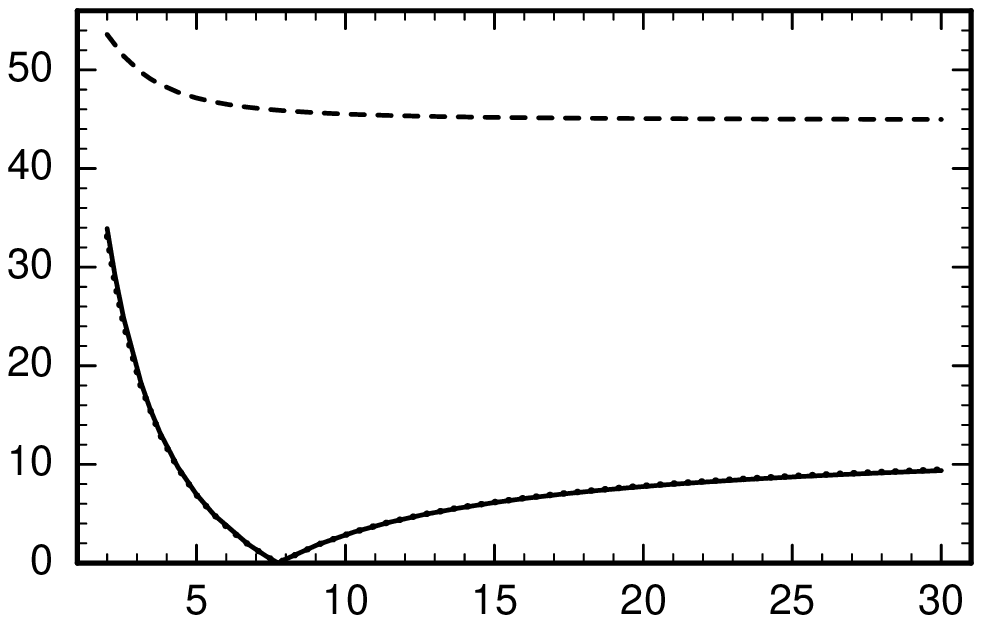,scale=0.9}}
\put(1.2,5){(c)}
\put(0.1,6.4){$|m_{\tilde{\chi}^0_1}|$ [GeV]}
\put(4.3,0.2){$\tan\beta$}
\end{picture}
\caption{Mass of the lightest neutralino in the MNSSM (solid), its
  upper bound according to Eq.~(\ref{8}) (dashed) and its approximate
  solution according to Eq.~(\ref{15}) (dotted) for $\lambda = 0.7$,
  $M_1 = 0.5 M_2$ and (a) $\tan\beta = 5$, $M_2 = 200$~GeV,
  (b) $\tan\beta = 5$, $\mu_{eff} = 200$~GeV,
  (c) $M_2 = \mu_{eff} = 200$~GeV.}
\end{figure}

As follows from Fig.~1 (a)--(c) the approximate solution (\ref{15}) describes 
the numerical one with relatively high accuracy even for
small $M_2\simeq\mu_{eff}\simeq 200\,\mbox{GeV}$. One can also see that 
the mass of the lightest neutralino may be very small or even
zero for large values $\lambda \sim 0.7$. 
This happens because the determinant of the neutralino mass matrix
(\ref{1}) is zero for a certain relation between 
the parameters (\ref{3}), namely, when
\be
M_1 M_2 \mu_{eff}\sin 2\beta=\tilde{M} M_Z^2\,.
\label{16}
\ee
The condition (\ref{16}) is fulfilled automatically when $M_1\sim M_2\to 0$. It means that the lightest neutralino mass always
vanishes when $M_1$ and $M_2$ go to zero. At the same time condition
(\ref{16}) can be satisfied at non--zero values of the soft gaugino
masses. This can be seen in Fig.~1 (a)--(c). Since in Fig.~1 (a) and
(c) we plot $|m_{\chi^0_1}|$ for non--zero values of the soft gaugino
masses the lightest neutralino mass vanishes only once. At the same
time in Fig.~1 (b) where we examine the dependence of $|m_{\chi^0_1}|$
on $M_2$ the mass of the lightest neutralino vanishes twice: once for
$M_2=0$ and once for a non--zero value of $M_2$ that obeys Eq.~(\ref{16}).
In the approximate solution (\ref{15}) the vanishing of the mass of the lightest neutralino corresponds to the vanishing of $C$, 
which is proportional to the determinant of the neutralino mass matrix (\ref{1}), i.e. 
$C=\ds\frac{\mbox{det}\,M_{\tilde{\chi}^0}}{(\mu_{eff}^2+\nu^2)(M_1+M_2)}$.

\begin{figure}[t]
\begin{picture}(16,8.5)
\put(0,0.4){\epsfig{file=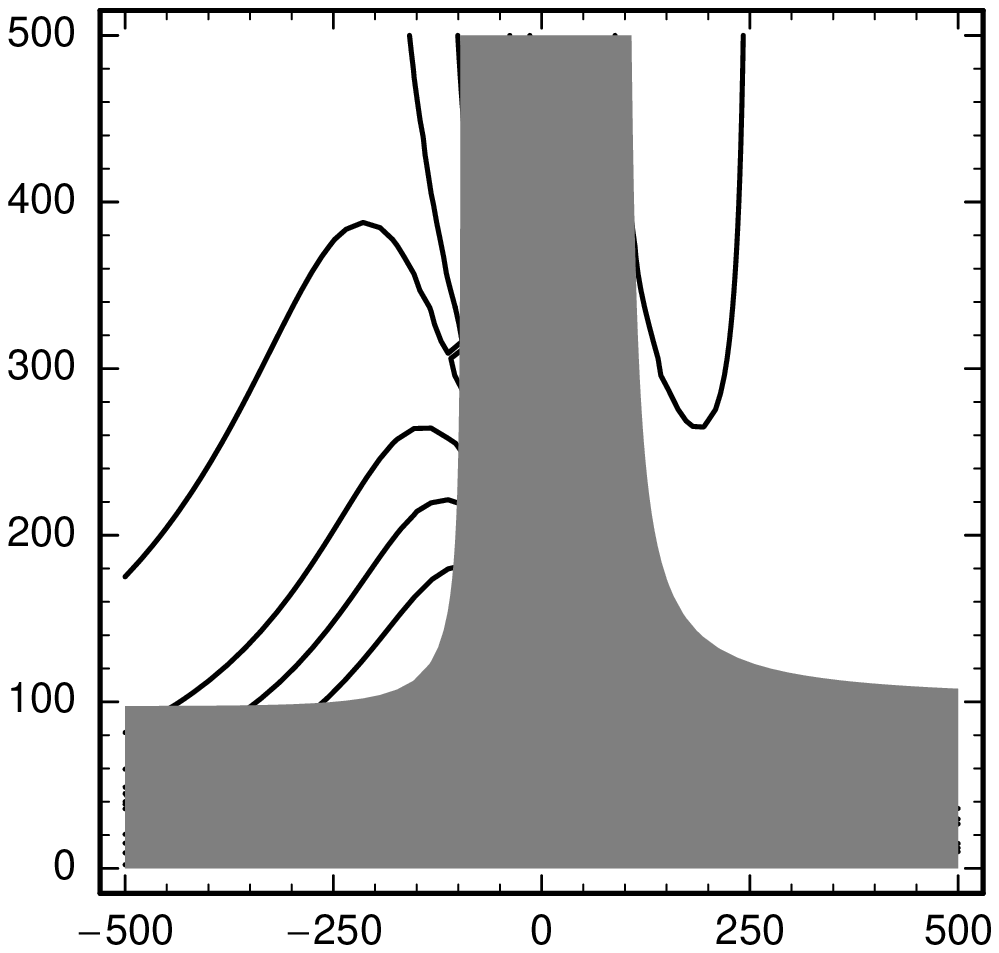,scale=0.75}}
\put(1.1,7.3){(a)}
\put(0.1,8){$M_2$ [GeV]}
\put(3.2,0.2){$\mu_{eff}$ [GeV]}
\put(2.6,6.3){\tiny $0.1$}
\put(3,4.75){\tiny $0.5$}
\put(3.25,4.15){\tiny $1$}
\put(3.25,3.6){\tiny $2$}
\put(5.75,7.4){\tiny $0.1$}
\put(2.7,1.8){$m_{\tilde{\chi}^\pm_1} < 100$ GeV}

\put(8.4,0.4){\epsfig{file=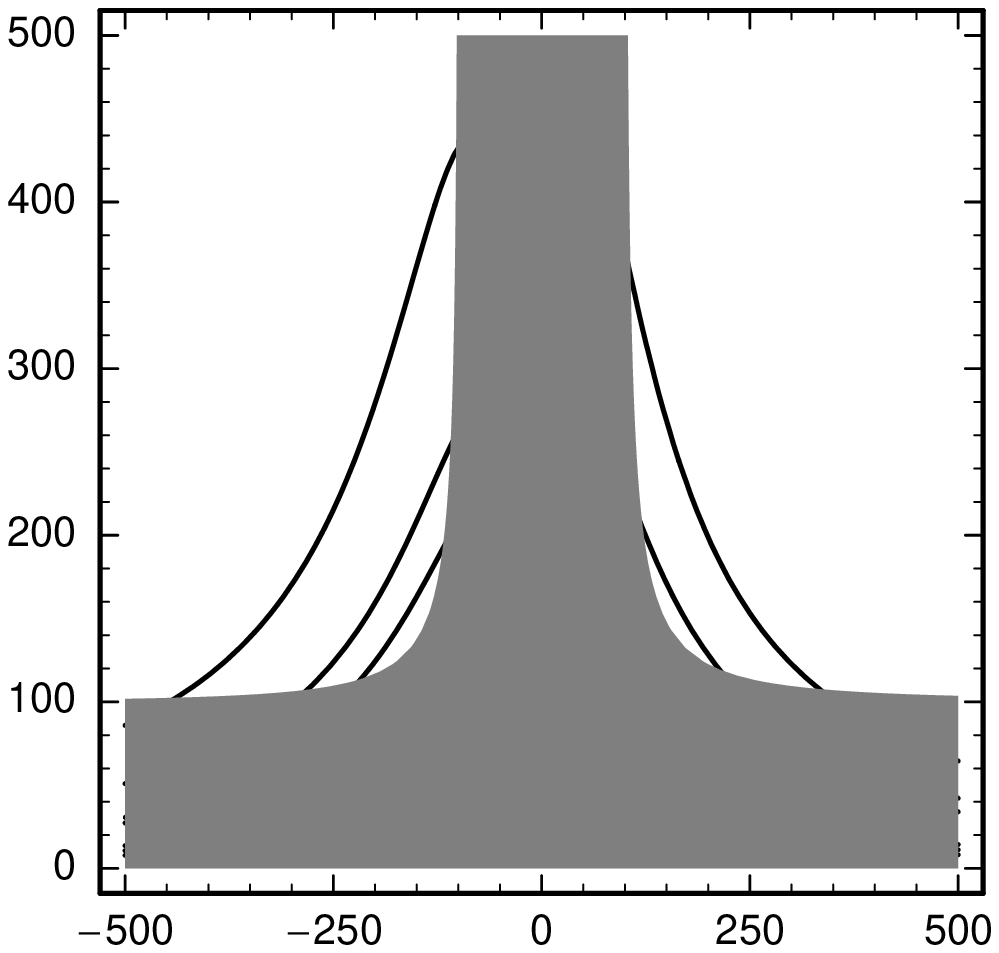,scale=0.75}}
\put(9.5,7.3){(b)}
\put(8.5,8){$M_2$ [GeV]}
\put(11.6,0.2){$\mu_{eff}$ [GeV]}
\put(11.5,6.75){\tiny $0.1$}
\put(11.45,4.5){\tiny $0.5$}
\put(11.65,3.75){\tiny $1$}
\put(13.35,3.8){\tiny $0.5$}
\put(13.25,5.9){\tiny $0.1$}
\put(11.1,1.8){$m_{\tilde{\chi}^\pm_1} < 100$ GeV}

\end{picture}
\caption{\label{fig:contour}Contours of the absolute value of the
difference between the mass of the lightest neutralino in the MNSSM
and its approximate solution (\ref{15}) in [GeV] for
$\lambda = 0.7$, $M_1 = 0.5 M_2$ and (a)
$\tan\beta = 5$, (b) $\tan\beta = 30$.
In the shaded region is $m_{\tilde{\chi}^\pm_1} < 100$~GeV.}
\end{figure}

Figure~\ref{fig:contour} shows the contours of the difference between
the exact and the approximate solution (\ref{15}) in the
$\mu_{eff}$-$M_2$ parameter plane. It can be seen that this difference
is smaller than 1~GeV in most of the phenomenologically allowed
parameter space, in large regions even smaller than 0.1~GeV.

Finally we would like to add that the two solutions of the reduced form of the characteristic equation (\ref{11})
describe with good accuracy not only the lightest neutralino mass but
also the mass of the second lightest 
one if the second lightest neutralino is considerably lighter than the
other states. Such a pattern of the neutralino 
spectrum is realised, for example, when $ M_1\ll M_2, \mu_{eff}$. Although it is rather difficult
to find any justification of this scenario within SUSY GUT or string inspired models it is not excluded by either 
LEP or Tevatron searches. If $\nu, M_Z\lesssim M_1$ in the considered limit then the mass of the second lightest
neutralino can be approximated by
\be
|m_{\chi^0_2}|\simeq\biggl|\ds\frac{M_1 M_2}{M_1+M_2}\biggr|\,.
\label{19}
\ee
In this case the lightest and the second lightest neutralino are
predominantly singlino and the superpartner of the
$U(1)_Y$ gauge boson.

\subsection{Approximate solution for decoupling limit}

With increasing effective $\mu$--term and soft gaugino masses the lightest neutralino mass decreases (see Fig.~1 (a)--(b)). 
From Fig.~1 (a)--(c) it becomes clear that the difference between the numerical and approximate solutions reduces when $\mu_{eff}$,
$M_1$ and $M_2$ grow. If either $\mu_{eff}$ or $M_1$ and $M_2$ are
much larger than $M_Z$, $B^2\gg C$ and 
the approximate solution for the lightest neutralino mass can be presented in a more simple form:
\be
|m_{\chi^0_1}|\simeq\frac{C}{B}\simeq\ds\frac{|\mu_{eff}|\nu^2\sin 2\beta}{\mu^2_{eff}+\nu^2}\,.
\label{17}
\ee
According to Eq.(\ref{17}) the mass of the lightest neutralino is inversely proportional to the effective $\mu$--term.
It vanishes when $\lambda$ tends to zero. In the limit $\lambda\to 0$ the equations for the extrema of the Higgs boson effective 
potential which determines the position of the physical vacuum imply
that the vacuum expectation value of the singlet field rises
as $M_Z/\lambda$. In other words the correct breakdown of electroweak symmetry breaking requires $\mu_{eff}$ to remain constant
when $\lambda$ goes to zero. As a result from Eq.~(\ref{17}) it follows that the mass of the lightest neutralino is proportional 
to $\lambda^2$ at small values of $\lambda$. At this point the approximate solution (\ref{17}) improves the theoretical 
restriction on the lightest neutralino mass derived in the previous section. This is because at small values of $\lambda$
the upper bound (\ref{9}) is proportional to $\lambda$.

From Eq.~(\ref{17}) one can also see that the mass of the lightest neutralino decreases when $\tan\beta$ grows.
The numerical results of our analysis summarised in Fig.~1 (a)--(c) confirm that $|m_{\chi^0_1}|$ becomes smaller
when $\tan\beta$ raises from $3$ to $10$. However, if $\tan\beta\gtrsim \zeta=\ds\frac{2 M_1 M_2 \mu_{eff}}{\tilde{M} M_Z^2}$,
Eq.~(\ref{17}) does not provide an appropriate description of the lightest neutralino mass. Indeed, in accordance with 
Eq.~(\ref{17}) the mass of the lightest neutralino vanishes at large values of $\tan\beta$ while Fig.~1 (c) demonstrates 
that $|m_{\chi^0_1}|$ approaches to some constant non--zero value with raising of $\tan\beta$. More accurate consideration
of the approximate solution (\ref{15}) allows to reproduce the asymptotic behaviour of the lightest neutralino mass at 
$\mu_{eff}, M_2, M_1\gg M_Z$ and at very large $\tan\beta\gg \zeta$. It is given by
\be
|m_{\chi^0_1}|\to \ds\frac{\nu^2 M_Z^2}{\mu^2+\nu^2}\biggl|\frac{\tilde{M}}{M_1 M_2}\biggr|\,.
\label{18}
\ee
So once again the approximate solution (\ref{15}) improves the theoretical restriction on the lightest neutralino 
mass because the upper limit (\ref{8})--(\ref{9}) on $|m_{\chi^0_1}|$ obtained before depends rather weakly on
$\tan\beta$.

\section{Conclusions}

In this letter we have examined the theoretical restrictions on the lightest neutralino mass within the
Minimal Non--minimal Supersymmetric Standard Model. In order to derive the appropriate upper bound
we consider the hermitian matrix $M_{\tilde{\chi}^0} M^{\dagger}_{\tilde{\chi}^0}$ where
$M_{\tilde{\chi}^0}$ is the neutralino mass matrix. The eigenvalues of
this matrix are the absolute values
of the neutralino masses squared. Therefore all eigenvalues of $M_{\tilde{\chi}^0} M^{\dagger}_{\tilde{\chi}^0}$
are positive definite. Using the theorem that the smallest diagonal
element of a hermitian matrix is always 
larger than the minimal eigenvalue of this matrix we establish an upper bound on the mass of the 
lightest neutralino $m_{\chi^0_1}$ in the MNSSM. The direct application of this theorem leads to the 
conclusion that $|m_{\chi^0_1}|$ has to be always less than $|\lambda|
v/\sqrt{2}$. A more stringent limit 
on the lightest neutralino mass can be obtained by applying an unitary transformation to the matrix
$M_{\tilde{\chi}^0} M^{\dagger}_{\tilde{\chi}^0}$. As a result we have
found that $|m_{\chi^0_1}|$
does not exceed $80-85\,\mbox{GeV}$. The corresponding upper bound depends rather strongly on 
the effective $\mu$--term $|\mu_{eff}|$
which is generated after the electroweak symmetry breaking. At large values
of $|\mu_{eff}|$ the upper limit on $|m_{\chi^0_1}|$ goes to zero so
that the mass interval 
of the lightest neutralino shrinks drastically.

Assuming that $|m_{\chi^0_1}|$ is considerably less than the masses of
the other neutralino states 
we have derived an approximate solution for the lightest neutralino mass. The obtained 
solution describes the numerical one with high accuracy in a large region of
the phenomenologically allowed part of the MNSSM parameter space. Our numerical analysis and analytic considerations
show that $m_{\chi^0_1}$ decreases with increasing $\tan\beta$  and
decreasing coupling $\lambda$. 
At small values of $\lambda$ the mass of the lightest neutralino is proportional to $\lambda^2$.
The lightest neutralino mass also decreases with increasing $\mu_{eff}$, $M_1$, and $M_2$.
We have argued that at large values of the effective $\mu$--term $m_{\chi^0_1}$ is
inversely proportional to $\mu_{eff}$. In the allowed part of the parameter space the 
lightest neutralino is predominantly singlino that makes its
direct observation at future colliders challenging. In forthcoming publications we plan to consider 
the potential discovery of such a neutralino at the LHC and ILC.

In summary,
the obtained theoretical restriction on the lightest neutralino mass
allows to discriminate the
MNSSM from other SUSY models where the mass of the lightest neutralino
is not limited from above. If no light neutralino is detected at future colliders the 
MNSSM will be ruled out.

\section*{Acknowledgements}

\vspace{-2mm} The authors would like to thank P.M.~Zerwas for his continual support
and encouragement. The authors are grateful to A.~Djouadi, J.~Kalinowski, D.~I.~Kazakov, S.~F.~King 
and P.~Langacker for valuable comments and remarks. RN would also like to thank 
E.~Boos, C.~Froggatt, V.~A.~Rubakov, D.~Sutherland and M.~I.~Vysotsky for fruitful discussions. 
RN acknowledge support from the SHEFC grant HR03020 SUPA 36878.

\end{document}